\documentstyle[12pt]{article}
\newcommand{\be}{\begin{equation}}
\newcommand{\ee}{\end{equation}}

\def\dfrac#1#2{{\displaystyle {#1 \over #2}}}
\def\dint{\displaystyle \int }
\def\dsum{\displaystyle \sum }

\begin{document}
\begin{titlepage}
\begin{center}
\Large\bf
 On Holomorphic Effective Actions of Hypermultiplets Coupled to
 External Gauge Superfields \\[1cm]
\large
          I.L. Buchbinder$^{\dag }$ and I.B. Samsonov$^{\ddag }$ \\[1cm]
{\small\it
$^{\dag }$ Department of Theoretical Physics, Tomsk State Pedagogical
University, Tomsk 634041, Russia \\
$^{\ddag }$ Department of Quantum Field Theory, Tomsk State University,
Tomsk 634050, Russia }\\[1cm]
                    Abstract
\end{center}
\begin{quotation}
We study the structure of holomorphic effective action for hypermultiplet
models interacting with background super Yang-Mills fields. A general form
of holomorphic effective action is found for hypermultiplet belonging to
arbitrary representation of any semisimple compact Lie group spontaneously
broken to its maximal abelian subgroup. The applications of obtained
results to hypermultiplets in fundamental and adjoint representations of
the $SU(n)$, $SO(n)$, $Sp(n)$ groups are considered.
\end{quotation}
\end{titlepage}
One of the remarkable properties of supersymmetric field theories is an
existence of two types of effective lagrangians, chiral and general or
holomorphic and non-holomorphic. General or non-holomorphic lagrangian
contributes to effective action in form of integral over full superspace
while chiral or holomorphic lagrangian contributes to effective action in
form of integral over chiral subspace. We point out that an appearance of
holomorphic corrections to effective action was firstly demonstrated in
refs \cite{West}-\cite{Petrov} (see also \cite{SuSy}, \cite{Petrov1}) for
$N=1$ SUSY and in refs \cite{Divecchia}, \cite{Seib88} for $N=2$ SUSY.

The modern interest to effective action in $N=2$ supersymmetric theories is
conditioned by the work \cite{Seib} where exact instanton contribution to
holomorphic effective potential has been found for $N=2$ $SU(2)$ super
Yang-Mills model. Generalization of this result to theories with the gauge
groups $SU(n)$ and $SO(n)$ have been given in refs \cite{SUn}, \cite{SOn}.

It is well known, a most simple and clear way of studying of effective
action in SUSY models is based on a formulation of these models in terms
of unconstrained superfields. Namely such a formulation provides a
possibility to preserve a manifest supersymmetry at all stages of quantum
calculations. For $N=2$ SUSY the corresponding formalism was developed in
refs \cite{harm} and was called harmonic superspace approach.

In this note we study a structure of holomorphic effective action for the
theory of hypermultiplet coupled to external super Yang-Mills field in
harmonic superspace approach. It is assumed that the hypermultiplet
belongs to arbitrary representation of any semisimple compact Lie group
and the external superfield lies in Cartan subalgebra of the corresponding
Lie algebra.

Holomorphic effective action for hypermultiplets coupled to abelian gauge
superfield has been obtained in harmonic superspace approach in
refs \cite{Buch1}, \cite{Buch1_} as follows
\be
\begin{array}{c}
\Gamma_{q,\omega }[V^{++}]=\dint d^4xd^4\theta {\cal F}_{q,\omega }(W)+c.c.,
 \\
{\cal F}_{q,\omega}(W)=
  -\dfrac{k_{q,\omega}}{64\pi ^2}W^2\log\dfrac{W^2}{\Lambda ^2},
\quad
k_{q,\omega}=\left\{
  \begin{array}{l}
    {\rm 1 \ for \ } q {\rm -hypermultiplet} \\
    {\rm 2 \ for \ } \omega {\rm -hypermultiplet}.
  \end{array}
  \right.
\end{array}
\label{e3}
\ee
Here $W$ is the strength of $N=2$ gauge prepotential $V^{++}$ \cite{harm},
\cite{Zupnik}
\be
W=-\int du(\bar D^{-})^2V^{++}(x,\theta ,u),\quad
\bar W=-\int du(D^{-})^2V^{++}(x,\theta ,u).
\label{e4}
\ee
The $q$- and $\omega$-hypermultiplets have been described in refs
\cite{harm}. Function ${\cal F}_{q,\omega}(W)$ is called the holomorphic
potential.

In this note we consider the $q$- and $\omega$-hypermultiplets $Q^{+}$
and $\Omega$ in an arbitrary representation of some semisimple compact
gauge group. Let $e_i$ $(i=1,\ldots ,m)$ be a basis in representation
space $\cal V$, $\dim {\cal V}=m$, and $T_i$ be the generators of the
representation forming a basis in Lie algebra $\cal L$. These
hypermultiplets and non-abelian gauge superfield $V^{++}$ can be written
in the forms
\be
Q^{+}=\sum^m_{i=1}q_i^{+}e_i,\quad
\breve Q^{+}=\sum^m_{i=1}\breve q_i^{+}e_i^{\dagger },\quad
\Omega =\sum^m_{i=1}\omega_ie_i,\quad
\breve \Omega =\sum^m_{i=1}\breve \omega_ie_i^{\dagger },
\label{e5}
\ee
\be
\quad V^{++}=\sum_kV_k^{++}T_k, \\
\label{e5_}
\ee
The classical actions of the hypermultiplets coupled to the external
superfield $V^{++}$ look like as follows
\be
S[Q^{+},\breve Q^{+},V^{++}] = \int d\zeta^{(-4)}\breve
Q^{+}(D^{++}+iV^{++})Q^{+}
\label{e6}
\ee
\be
S[\Omega ,\breve\Omega ,V^{++}] = \int d\zeta^{(-4)}
(D^{++}\breve\Omega -i\breve\Omega V^{++})(D^{++}+iV^{++})\Omega .
\label{e7}
\ee
We use the denotations like in refs \cite{Buch1}, \cite{Buch1_}. Manifest
form of the terms $V^{++}Q^{+}$, $V^{++}\Omega$ depends on
representation, e.g., in the adjoint representation these terms are
understood as the
commutators. The expression $\breve Q^{+}D^{++}Q^{+}$
means $\breve q^{+}_iD^{++}q^{+}_j(e_i,e_j)$ where
$(e_i,e_j)$ is a scalar product of vectors in a representation space.

Further we assume that the external superfield $V^{++}$ lies in the Cartan
subalgebra $\cal H$ of gauge algebra $\cal L$. Therefore
\be
V^{++}=\sum^l_{k=1} V^{++}_k h_k,
\label{e8}
\ee
where $l={\rm rank}\cal L$ and $\{h_1,\ldots,h_l\}$ is a basis in $\cal H$.

The main idea of subsequent consideration is to reduce a calculation of
effective actions for the models (\ref{e6}), (\ref{e7}) to calculation of
effective actions for hypermultiplets in abelian external gauge superfield
and use the result (\ref{e3}). To do that, we "diagonalise" the
actions (\ref{e6}), (\ref{e7}) and obtain several "one-dimensional"
hypermultiplets in abelian gauge superfields. The realization of this idea
is based on use of notion of the weights of Lie group representation.

The weight subspaces ${\cal V}^\lambda$ of representation space $\cal V$
are defined as follows \cite{Naimark}
\be
{\cal V}^{\lambda }=\{v\in {\cal V}\ | \ hv=\lambda (h)v, \ \forall h\in
{\cal H} \}.
\label{e9}
\ee
Let the equation (\ref{e9}) have $m=\dim \cal V$ roots (weights)
$\lambda_i$ (some of which may be equal) and corresponding eigenvectors
$e_i$ be orthogonal
\be
(e_i,e_j)=\delta _{ij}.
\label{e10}
\ee
We choose vectors $e_i$ as the basis in representation space and
expand hypermultiplets (\ref{e5}) in this basis.
With the help of relation (\ref{e9}) it is easy to find the
action of gauge superfield $V^{++}$ on the hypermultiplets
\be
V^{++}Q^{+}=\sum^m_{i=1}q^{+}_i\lambda_i(V^{++})e_i, \quad
V^{++}\Omega =\sum^m_{i=1}\omega _i\lambda _i(V^{++})e_i.
\label{e13}
\ee
It is easy to see that functions $\lambda_i(V^{++})$ are linear
$\lambda_i(V^{++})=\lambda_i(h_k)V^{++}_k$, so the weights
$\lambda_i(V^{++})$ have $U(1)$ charge $+2$, the same as the charge of
$V^{++}$. Substituting the hypermultiplets (\ref{e5}) into
actions (\ref{e6}), (\ref{e7}) and using eqs (\ref{e10}), (\ref{e13})
we obtain
\be
S[\breve Q^{+},Q^{+},V^{++}]=\int d\zeta^{(-4)}\sum^m_{i=1}
 \breve q^{+}_i(D^{++}+i\lambda_i(V^{++}))q^{+}_i,
\label{e14}
\ee
\be
S[\breve \Omega ,\Omega ,V^{++}]=\int d\zeta^{(-4)}\sum^m_{i=1}
 (D^{++}-i\lambda_i(V^{++}))\breve\omega_i(D^{++}+i\lambda_i(V^{++}))\omega_i.
\label{e15}
\ee
The hypermultiplets labeled by different indices in (\ref{e14}),
(\ref{e15}) do not mix between themselves, therefore their classical
actions are the sums of actions of non-coupled "one-dimensional"
hypermultiplets. Hence, holomorphic effective actions of models
(\ref{e14}), (\ref{e15}) are represented as a sum of independent
"one-dimensional" effective actions (\ref{e3}) over all weights of
representation
\be
\Gamma_{q,\omega}[V^{++}]=\dint d^4xd^4\theta {\cal F}_{q,\omega }(W)+c.c.,
\quad
{\cal F}_{q,\omega}(W)=
 -\frac{k_{q,\omega}}{64\pi^2}\dsum^m_{i=1} W^2_i\log\frac{W^2_i}{\Lambda^2},
\label{e16}
\ee
where $W_i$ are constructed of weights $\lambda_i(V^{++})$ according to
(\ref{e4}).

Thus we have obtained the effective actions of $q$- and
$\omega$-hypermultiplets in arbitrary representation of any semisimple
gauge group (\ref{e16}), they are expressed via weights of the
representation. It is well known that any irreducible representation is
completely determined by its highest weight $\alpha $ and all other
weights of representation are expressed of $\alpha$ (see \cite{Naimark})
\be
\lambda (h)=\alpha (h)-m_1\beta_1(h)-m_2\beta_2(h)-\ldots -m_k\beta_k(h),
\ h\in {\cal H},
\label{e17}
\ee
where $\beta_i(h)$ are simple roots of algebra $\cal L$, $m_1,\ldots ,m_k$
are non-negative integers. Hence, we see the effective action
(\ref{e16}) is actually determined by highest weight $\alpha $.

Now we are ready to write down holomorphic effective actions of $q$- and
$\omega$-hypermultiplets in the fundamental and adjoint representations of
$SU(n)$, $SO(n)$, $Sp(n)$ groups.
To do this we have to fix Cartan subalgebra of each group, find all
weights of corresponding representation, and substitute these weights into
(\ref{e16}). Using the weights of representations and structure of the
Cartan subalgebras given in the Appendix one obtain holomorphic
effective actions of hypermultiplets on the base of eq. (\ref{e16}). Below
we give a list of corresponding holomorphic effective potentials:\\
1. Fundamental representation
\be
{\cal F}^{SU(n)}_{q,\omega}(W)=
-\frac{k_{q,\omega }}{64\pi^2}\sum^n_{i=1}W^2_i\log\frac{W^2_i}{\Lambda^2},
\label{e18}
\ee
\be
{\cal F}^{SO(2n)}_{q,\omega}(W)={\cal F}^{SO(2n+1)}_{q,\omega}(W)=
{\cal F}^{Sp(n)}_{q,\omega}(W)=
-\frac{k_{q,\omega }}{32\pi^2}\sum^n_{i=1}W^2_i\log\frac{W^2_i}{\Lambda^2};
\label{e19}
\ee
2. Adjoint representation
\be
{\cal F}^{SU(n)}_{q,\omega}(W)=
-\frac{k_{q,\omega }}{32\pi^2}\sum^n_{i<j}(W_i-W_j)^2
\log\frac{(W_i-W_j)^2}{\Lambda^2},
\label{e20}
\ee
\be
{\cal F}^{SO(2n)}_{q,\omega}(W) =
 -\frac{k_{q,\omega}}{64\pi^2}\dsum_{i<j}^n\left[ (W_i-W_j)^2
\log \frac{(W_i-W_j)^2}{\Lambda ^2}+(W_i+W_j)^2
\log \frac{(W_i+W_j)^2}{\Lambda ^2}\right],
\label{e21}
\ee
\be
\begin{array}{ll}
{\cal F}^{SO(2n+1)}_{q,\omega}(W)=&
   -\dfrac{k_{q,\omega}}{64\pi^2} \left[\sum\limits ^n_{i<j}(W_i-W_j)\log
\frac{(W_i-W_j)^2}{\Lambda ^2} \right. \\
 &+\left. \sum\limits ^n_{i<j}
(W_i+W_j)^2\log\frac{(W_i+W_j)^2}{\Lambda ^2}
 +\sum\limits ^n_{l=1}W_l^2\log \frac{W_l^2}{\Lambda ^2}\right].
\end{array}
\label{e22}
\ee
\be
{\cal F}^{Sp(n)}_{q,\omega}(W)=-\dfrac{k_{q,\omega}}{64\pi ^2}
 \left[ \sum\limits_{i<j}^n  (W_i-W_j)^2
\log \frac{(W_i-W_j)^2}{\Lambda ^2}+
 \sum\limits_{i\leq j}^n (W_i+W_j)^2
 \log \frac{(W_i+W_j)^2}{\Lambda^2} \right] ,
\label{e23}
\ee
where $W_i$ are $N=2$ SYM strengths constructed of $V^{++}_i$ by the rule
(\ref{e4}).

Some of these results were obtained earlier in refs
\cite{SUn,SOn,Buch2,Buch3,Ivanov} by different methods. In our approach
the holomorphic effective actions are easily obtained on the base of
single method from the general expression (\ref{e16}).

To conclude, in the present paper we have considered the
hypermultiplet models coupled to $N=2$ nonabelian gauge
superfield using the harmonic superspace formalism. We obtained the
holomorphic effective actions of these models for the general case when
hypermultiplets belong to arbitrary representation of any semisimple gauge

group and gauge superfield lies in Cartan subalgebra of the gauge algebra.
Then we applied these results to find the effective actions of
$q$- and $\omega$-hypermultiplets in fundamental and adjoint
representations of $SU(n)$, $SO(n)$, $Sp(n)$ groups.

The authors are very grateful to E.A. Ivanov, S.M. Kuzenko,
B.M. Zupnik for useful comments. The work of I.L.B. was
supported in part by RFBR, project No 99-02-16017; RFBR-DFG, project No
96-02-04022; INTAS, project INTAS-96-0308; GRACENAS, project No 97-6.2-34.
\\[0.5cm]
{\Large\bf Appendix}
\nopagebreak \\
We formulate a list of useful properties of considered
algebras such as the structures of Cartan subalgebras and weights of
fundamental and adjoint representations. These properties have been used
to obtain the effective potentials (\ref{e18})-(\ref{e23}) (see the details
in refs \cite{Naimark}, \cite{Barut}).

1. Fundamental representation.\nopagebreak

The basis in representation space $e_i$, Cartan subalgebra-valued
superfield $V^{++}$ and corresponding weights of fundamental representation
have been chosen in the form

\begin{tabular}{ll}
\hline\hline
$SU(n)$  & $(e_i)_k = \delta_{ik}\quad i,k=1,\ldots ,n$ \\
& $V^{++}={\rm diag}(V^{++}_1,V^{++}_2,\ldots ,V^{++}_n),\ \
\sum\limits^n_{i=1}V^{++}_i=0 $\\
& $\lambda_i(V^{++})=V^{++}_i$;
\\
\hline
$SO(2n)$ &
$(e_i)_j=\frac1{\sqrt2}(i\delta_{2i-1,j}+\delta_{2i,j}), \quad
(e_{-i})_j=\frac1{\sqrt2}(\delta_{2i-1,j}+i\delta_{2i,j}), \
i,j=1,\ldots ,n $\\
& $V^{++}={\rm diag}(V^{++}_1I,\ldots ,V^{++}_nI), \quad
I= \left(
\begin{array}{cc}
0 & i \\
-i & 0
\end{array}
\right) $\\
& $\lambda_i(V^{++})=V^{++}_i$; \\
\hline
$SO(2n+1)$ &
$ (e_i)_j=\frac1{\sqrt2}(i\delta_{2i-1,j}+\delta_{2i,j}), \
(e_{-i})_j=\frac1{\sqrt2}(\delta_{2i-1,j}+i\delta_{2i,j}), $\\
&\quad $(e_0)_i=\delta_{2n+1,i},\ i,j=1,\ldots ,n $ \\
& $V^{++}={\rm diag}(V^{++}_1I,\ldots ,V^{++}_nI,0), \quad
I=\left(
\begin{array}{cc}
0 & i \\
-i & 0
\end{array}
\right) $\\
& $\lambda_i(V^{++})=V^{++}_i\ (i\ne 0),\ \lambda_0(V^{++})=0 $;
\\
\hline
$Sp(n)$ &
$(e_i)_k=\delta _{ik}, \quad i,k= 1,\ldots ,2n $\\
& $V^{++}={\rm diag}\{V^{++}_1,\ldots ,V^{++}_n,
          -V^{++}_1,\ldots ,-V^{++}_n  \} $\\
& $ \lambda_i(V^{++})=V^{++}_i $. \\
\hline\hline
\end{tabular}

2. Adjoint representation.

In adjoint representation the weight subspaces coincide with the root
subspaces in a gauge algebra and the weights of the representation are the
roots of algebra. Therefore we have to fix Cartan-Weyl basis in each
considered algebra and write down its roots.

\begin{tabular}{ll}
\hline\hline
$SU(n)$ &
Cartan-Weyl basis:
$\{E_{\lambda_{ij}},\ H_m \} \quad i\ne j,\,i,j=1,\ldots ,n, \
 m=1,\ldots ,n-1 $\\
&$\quad (E_{\lambda_{ij}})_{kl}=\delta_{ik}\delta_{jl}, \quad
 (H_m)_{kl} = \delta_{mk}\delta_{ml}-\delta_{nk}\delta_{nl} $\\
&$[V^{++},E_{\lambda_{ij}}]=\lambda_{ij}(V^{++})E_{\lambda_{ij}}, \quad
\lambda _{ij}(V^{++})=V_i^{++}-V_j^{++}\, (i\ne j); $\\
\hline
$SO(2n)$ &
Cartan-Weyl basis:
$\{E_{\lambda _{ij}},E_{-\lambda _{ij}},H_k\},
 \quad i\ne j,\ i,j,k=1,\ldots ,n $\\
&$\quad E_{\lambda _{ij}}=\frac 1{\sqrt{8}}(i\tilde e_{2i-1,2j-1}+
 \tilde e_{2i-1,2j}-\tilde e_{2i,2j-1}+i\tilde e_{2i,2j}),\quad i<j, $\\
&$\quad E_{\lambda _{ij}}=\frac 1{\sqrt{8}}(i\tilde e_{2i-1,2j-1}-
 \tilde e_{2i-1,2j}-\tilde e_{2i,2j-1}-i\tilde e_{2i,2j}),\quad i>j, $\\
&$\quad H_k=\frac{i}{\sqrt{2}}\tilde e_{2k-1,2k},\quad
 \tilde e_{ij}=\delta _{ik}\delta _{jl}-\delta _{jk}\delta _{il}, $\\
&$\quad E_{-\lambda _{ij}}=(E_{\lambda _{ij}})^{\dagger },\quad
 (H_k)^{\dagger }=H_k $\\
&$[V^{++},E_{\pm\lambda_{ij}}]=\pm\lambda_{ij}(V^{++})E_{\pm\lambda_{ij}},
\quad
\lambda _{ij}(V^{++})=\left\{
\begin{array}{ll}
\frac 1{\sqrt{2}}(V_i^{++}-V_j^{++}) & i<j \\
\frac 1{\sqrt{2}}(V_i^{++}+V_j^{++}) & i>j;
\end{array}
\right. $
\\ \hline
$SO(2n+1)$ &
Cartan-Weyl basis:
$\{E_{\lambda _{ij}},E_{-\lambda _{ij}},E_{\lambda _k},E_{-\lambda _k},
H_k \},\quad i\ne j,\ i,j,k=1,\ldots ,n $\\
&\quad $H_k$, $E_{\lambda_{ij}}$, $\lambda_{ij}$ are the same as in $SO(2n)$ \\
&$\quad
E_{\lambda _k}=\frac 1{\sqrt{2}}(i\tilde e_{2k-1,2n+1}-\tilde e_{2k,2n+1}),
\quad E_{-\lambda _k}=(E_{\lambda _k})^{\dagger },\quad k=1,\ldots ,n $\\
&$[V^{++},E_{\pm\lambda_k}]=\pm\lambda_k(V^{++})E_{\pm\lambda_k},\quad
\lambda_k(V^{++})=\frac1{\sqrt 2}V^{++}_k; $\\
\hline
$Sp(n)$ &
Cartan-Weyl basis:
$\{F_{kl},G_{ij},\tilde G_{ij} \},
 \quad 1\le i\le j\le n,\ k,l=1,\ldots ,n, $\\
&$\quad F_{kl}=\frac1{\sqrt{2}}(e_{kl}-e_{l+n,k+n}), \quad
 (e_{ij})_{kl}=\delta_{ik}\delta_{jl} $\\
&$\quad G_{ij}=\eta_{ij}(e_{i+n,j}+e_{j+n,i}),\quad
 \tilde G_{ij}=\eta_{ij}(e_{i,j+n}+e_{j,i+n}), $\\
&$[V^{++},F_{ij}]=\alpha_{ij}(V^{++})F_{ij}, \quad
 \alpha _{ij}(V^{++})=\frac1{\sqrt{2}}(V^{++}_i-V^{++}_j), $\\
&$[V^{++},G_{ij}]=\beta_{ij}(V^{++})G_{ij}, \quad
 \beta _{ij}(V^{++})=\frac1{\sqrt{2}}(V^{++}_i+V^{++}_j), $\\
&$[V^{++},\tilde G_{ij}]=-\beta_{ij}(V^{++})\tilde G_{ij}.$\\
\hline\hline
\end{tabular}

We point out, the structure of effective potentials
(\ref{e19})-(\ref{e23}) is defined by the choice of basis vectors given
above and the coefficients in effective potentials depend on the
normalisation of these vectors.


\begin{thebibliography}{99}
\bibitem{West} P. West, Phys. Lett. {\bf B258}, 357, 1991.
\bibitem{Jack} I. Jack, D.R.T. Jones, P. West, Phys. Lett. {\bf B258},
         382, 1991.
\bibitem{Shif} M.A. Shiffman, A.I. Vainstein, Nucl. Phys. {\bf B359}, 571,
         1991.
\bibitem{Petrov} I.L. Buchbinder, S.M. Kuzenko, A.Yu. Petrov, Phys. Lett.
         {\bf B321}, 372, 1994.
\bibitem{SuSy} I.L. Buchbinder, S.M. Kuzenko. {\it Ideas and Methods
        of Supersymmetry and Supergravity}, IOP Publishing, Bristol and
        Philodelphia, 1995; Revised edition, 1998.
\bibitem{Petrov1} I.L. Buchbinder, A.Yu. Petrov, Phys. Lett. {\bf B416},
        209, 1999.
\bibitem{Divecchia} P. Di Vecchia, R. Musto, F. Nicodemi, R. Pettorino,
        Nucl. Phys. {\bf B252}, 635,1985.
\bibitem{Seib88} N. Seiberg, Phys. Lett. {\bf B206}, 75, 1988.
\bibitem{Seib} N. Seiberg, E. Witten, Nucl. Phys. {\bf B426}, 19, 1994.
\bibitem{SUn}
        A. Klemm, W. Lerche, S. Yankielowicz, S.Theisen, Phys. Lett.
        {\bf B344}, 169, 1995;
        Int. J. Mod. Phys. {\bf A11}, 1929, 1996;
        P.C. Argyres, A.E. Ferrari, Phys. Rev. Lett {\bf 74}, 3931, 1995;
        M.R. Douglas, S.H. Shenker, Nucl. Phys. {\bf B447}, 271, 1995;
        A. Hannany, Y. Oz, Nucl. Phys. {\bf B452}, 283, 1995;
        S. Naik, Nucl. Phys. {\bf B538}, 137, 1999.
\bibitem{SOn}
        U.H. Danielsson, B. Sundborg, Phys. Lett. {\bf B358}, 273, 1995;
        A. Brandhuber and K. Landsteiner, Phys. Lett. {\bf B358}, 73, 1995;
        A. Hanany, Nucl. Phys. {\bf B466}, 85, 1996.
\bibitem{harm} A. Galperin, E. Ivanov, S. Kalitzin,  V. Ogievetsky, E.
        Sokatchev, Class. Quant. Grav. {\bf 1}, 469, 1984;
        A. Galperin, E. Ivanov, V. Ogievetsky, E. Sokatchev,
        Class. Quant. Grav. {\bf 2}, 601; 617 1985.
\bibitem{Buch1} I.L. Buchbinder, E.I. Buchbinder, E.A. Ivanov, S.M.
        Kuzenko, B.A. Ovrut, Phys. Lett. {\bf B412}, 309, 1997.
\bibitem{Buch1_} E.I. Buchbinder, I.L. Buchbinder, E.A. Ivanov and
        S.M. Kuzenko, Mod. Phys. Lett. {\bf A13}, 1071, 1998.
\bibitem{Zupnik} B.M. Zupnik, Phys. Lett. {\bf B183}, 175, 1978.
\bibitem{Buch2} I.L. Buchbinder, E.I. Buchbinder, S.M. Kuzenko, B.A.
        Ovrut, Phys. Lett. {\bf B417}, 61, 1998.
\bibitem{Buch3} I. Buchbinder, S. Kuzenko, B. Ovrut, {\it Covariant
        Harmonic Supergraphity for $N=2$ Super Yang-Mills Theories.} Proc.
        of Int.  Seminar "Supersymmetries and Quantum Symmetries", ed. by
        J. Wess and E.A. Ivanov, Dubna, 1977, Springer, 1999, pp. 20-36.
\bibitem{Ivanov} S. Eremin, E. Ivanov, {\it Holomorphic Effective
        Action of $N=2$ SYM Theory from Harmonic Superspace with Central
        Charges}, hep-th/9908054.
\bibitem{Naimark} M.A. Naimark, {\it Theory of Group Representations}.
        Nauka, 1976 (in Russian).
\bibitem{Barut} A.O. Barut, R. Raczka, Theory of Group Representations and
        Applications, PWN -- Polish Scientific Publishers, Warszawa, 1977.
\end{thebibliography}
\end{document}